\documentclass[final]{cvpr}

\usepackage{times}
\usepackage{epsfig}
\usepackage{graphicx}
\usepackage{amsmath}
\usepackage{amssymb}

\usepackage[pagebackref=true,breaklinks=true,colorlinks,bookmarks=false]{hyperref}

\def\cvprPaperID{****} 
\def\confYear{CVPR 2021}

\begin{document}

\title{Implementing a Detection System for COVID-19 based on \\ Lung Ultrasound Imaging and Deep Learning}

\author{
Carlos Rojas-Azabache$^1$, Karen Vilca-Janampa$^1$, Renzo Guerrero-Huayta$^2$, Dennis Núñez-Fernández$^2$ \\
$^1$ Universidad Nacional Mayor de San Marcos \\
$^2$ Universidad Nacional de Ingeniería \\
Lima, Peru \\
{\tt\small \{carlos.rojas6,10140251\}@unmsm.pe,\{rguerreroh,dnunezf\}@uni.pe}
}

\maketitle

\begin{abstract}
The COVID-19 pandemic started in China in December 2019 and quickly spread to several countries. The consequences of this pandemic are incalculable, causing the death of millions of people and damaging the global economy. To achieve large-scale control of this pandemic, fast tools for detection and treatment of patients are needed. Thus, the demand for alternative tools for the diagnosis of COVID-19 has increased dramatically since accurated and automated tools are not available. In this paper we present the ongoing work on a system for COVID-19 detection using ultrasound imaging and using Deep Learning techniques. Furthermore, such a system is implemented on a Raspberry Pi to make it portable and easy to use in remote regions without an Internet connection.
\end{abstract}


\section{Introduction}

The COVID-19 pandemic originated in Wuhan, China, in December 2019 and since that time it has resulted in 176 million infections and the deaths of more than 3.79 million people (as of June 2021) \cite{1}. COVID-19 has now spread to various parts of the world and variants have even appeared in different geographic areas, which still continue having negative consequences. The most detrimental effects have been the death of millions of people, the collateral effects on those who have recovered, the damage to the world economy, and the increase in the level of global poverty. To counteract these effects and efficiently manage the pandemic, early detection of COVID-19 and isolation measures must be maintained \cite{2}. However, most early detection tools lack practicality, are expensive, and cannot be used in remote regions. The progress of Deep Learning methodologies has allowed us in recent years to diagnose diseases with high accuracy, thus tools have been developed for the prediction of COVID-19 based on Deep Learning.

\section{Previous works}

In the last year, and during the course of the pandemic, different solutions have been proposed for the early diagnosis of COVID-19 based on machine learning, deep learning and artificial intelligence techniques among others. Most of these solutions make use of medical images taken from ultrasound scans, computed tomography scans and chest X-rays. However, each methodology and imaging source used has had advantages and disadvantages. We will evaluate the most interesting proposals below.

In \cite{5}, the authors use a novel deep neural network, which is derived from Spatial Transformation Networks, and also make use of images taken with ultrasound scanners. This predicts the quantification of the severity of COVID-19 infection and provides the spatial localization of pathological features in ultrasound images. In \cite{3}, which was developed in 2020, X-ray, ultrasound and computed tomography images are used, together with a convolutional neural network based on the VGG-19 model and using transfer learning, reaching an accuracy of 84\% for computed tomography, 86\% for chest X-ray and 100\% for ultrasound. 

In another work of the same year, \cite{12}, a COVID-19 detection system is proposed that uses X-ray imaging together with deep learning techniques with relatively high accuracy. However, the use of X-rays makes the system costly, impractical, and it cannot be used repeatedly on a person to follow up their recovery due to its radiation. In another recent work, \cite{4}, they make use of about 1000 ultrasound images of three classes of patients: with COVID-19, with bacterial pneumonia and healthy patients. Thus, they perform the training of a convolutional neural network, which they called POCOVID-Net, and which gave a sensitivity of 96\% and a specificity of 79\%. However, the low number of images considered may not provide a high generalization capability.

\section{Ultrsound imaging}

There are different imaging sources (ultrasonography, chest radiography and computed tomography) that can be used to predict whether a person has COVID-19 or is healthy. However, the risks and benefits of each type of imaging source will depend on the individual patient and the stage of disease progression.

Currently, the preferred imaging source for COVID-19 pneumonia is computed tomography (CT) since it is characterized by ground-glass opacity (GGO) abnormalities early in the disease, followed by the crazy paving pattern and, finally, consolidation in the later stage of the disease \cite{6} \cite{7}. Although CT scans are very useful, their cost is very high and their availability in health centers is very limited, and even more so in remote health centers. In addition to this, the process of sterilization and cleaning of the CT scanner causes delays in the care of other patients.

Another popular method used to detect COVID-19 cases is chest X-ray, but several requirements must be met for its use. To be able to use, it requires trained personnel who have a license to use radiation, a good maintenance service of the equipment is needed, and certain requirements are required such as permissions because radiation is used, it is also necessary that the place where it is used meets with various regulations. We must also emphasize that this procedure is not used for detection, it is used only for control.

Due to the advantages of ultrasound imaging mentioned above, this type of imaging source will be used in this project. More precisely, we will employ lung ultrasound images due it is a low-cost and highly available method that does not use ionizing radiation

\section{Overview of the system}

As objectives of this work we have the construction of a portable system that uses lung ultrasound images for the automatic detection of COVI-19 and that can explain such detection. For this purpose we will use convolutional neural networks to perform the classification and segmentation tasks. The final model will be used on a Raspberry Pi embedded platform to perform classification and segmentation on new images. Due to the relatively high processing capacity needed, a Movidus neural co-processor will be used, which will allow us to have a fast response, almost in real time and without the need of internet access since the inference process is performed in the embedded device. 

\begin{figure}[h]
\vskip 0.1in
\begin{center}
\includegraphics[width=0.7\linewidth]{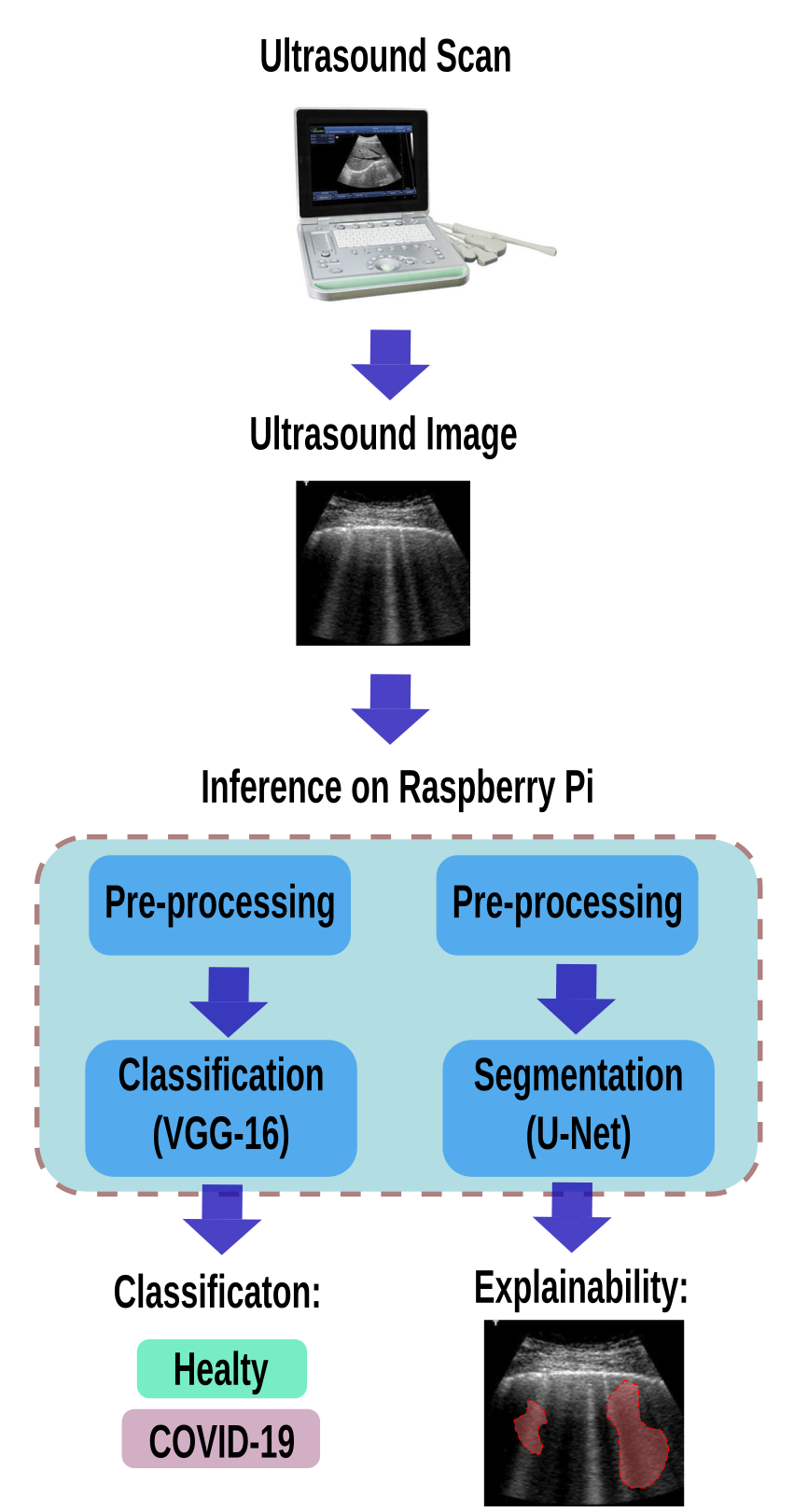}
\end{center}
\caption{Diagram of the proposed COVID-19 detection system}
\label{overview}
\vskip 0.1in
\end{figure}

In summary, see Fig.~\ref{overview}, the images are extracted with the Raspberry Pi from an ultrasound scanner, then these images are pre-processed and fed to a classification neural network (VGG-19) and a segmentation neural network (U-Net). Once the inference has been carried out, we will be able to visualize the classification and the areas where the prediction was made.

\section{CovidSET-Peru}

The dataset used for this project was collected in different private clinics in the city of Lima, and extracted with different brands of portable ultrasound scanners. The identification of the individuals were completely eliminated by the specialists of the clinics and only the ultrasound images without raw data were provided to us. We decided to name it "CovidSET-Peru" and one of the final intentions of this project is to leave it free in a repository so that researchers, developers and interested people can use it freely in future projects. We are confident that this data set will be useful for implementing better COVID-19 detection systems.

Currently, We are still collecting the images, organizing them and performing the manual classification and segmentation by specialists. Initially, we planned to have 50 videos, from which we will extract a total of 750 images for COVID-19 cases and 750 images for healthy cases. The extraction of the images was based on frame extraction. As more images become available, this dataset will be increased in future versions. The images are presented in 512x512 pixels in grayscale format.

\section{Proposed CNN architectures}

Before using the images as feed to the convolutional neural networks, a scaling to 224x224 pixels and a conversion to grayscale will be performed. In addition, for training, a data augmentation will be used, performing horizontal and vertical displacements, reflection in both axes and noise addition, thus obtaining at least 10 times the initial amount of images \cite{9}. 

\begin{figure}[h]
\vskip 0.1in
\begin{center}
\includegraphics[width=1.0\linewidth]{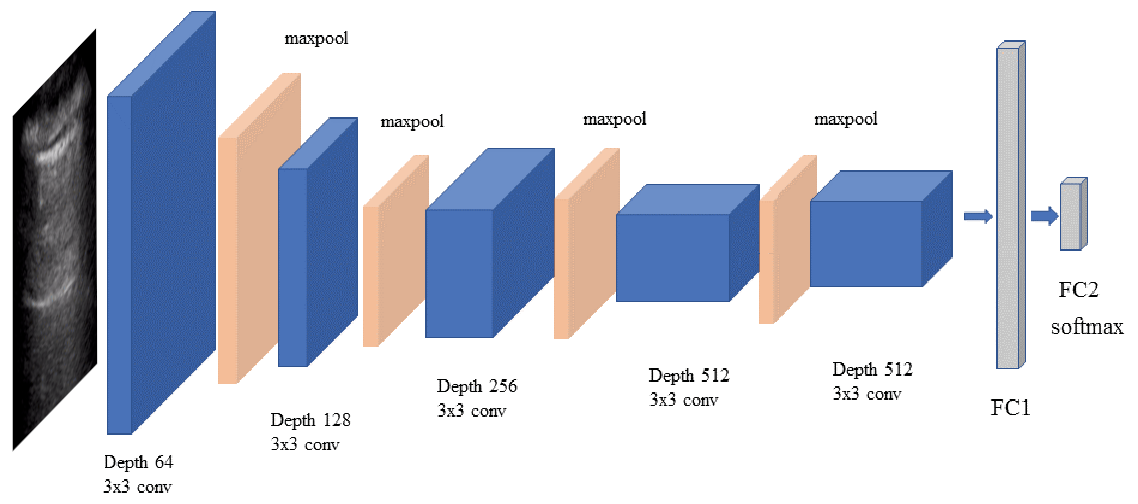}
\end{center}
\caption{VGG-16 based architecture for classification}
\label{vgg-16}
\vskip 0.1in
\end{figure}

For the classification task, we have been taken into account and studied the architectures from previous works. Thus, the models used for the classification of chest X-ray images such as VGG-16 \cite{10} and COVIDX-Net, which is based on different architectures (VGG-19, DCNN, Dense-Net 201, InceptionV2, Resnet-101, Inception V3, MobileNet V2 and Xception) \cite{11} were analyzed. From these it was concluded that the most suitable model is the VGG-16 architecture. The proposed architecture will have the structure shown in Fig.~\ref{vgg-16}. For a more optimal model construction, we will use transfer learning, employing the POCOVID-Net convolutional network \cite{4}. In this way, the full connection layers will be trained to classify the lung ultrasound images found in the proposed CovidSET-Peru dataset.

\begin{figure}[h]
\vskip 0.1in
\begin{center}
\includegraphics[width=1.0\linewidth]{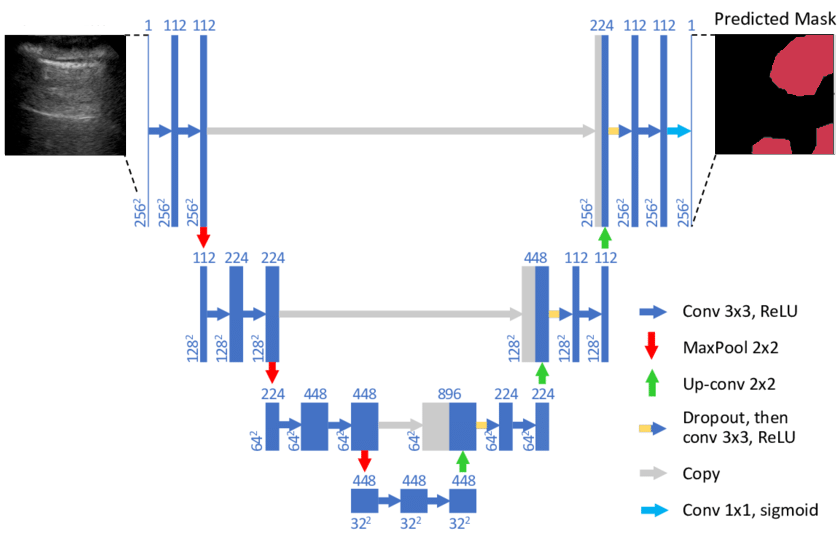}
\end{center}
\caption{U-Net based architecture for segmentation}
\label{u-net}
\vskip 0.1in
\end{figure}

For the segmentation task, we make use of the U-Net architecture and some hyper parameters were modified in order to work with our collected dataset and to produce an accurate segmentation. As Fig.~\ref{u-net} shows, the U-Net consists of a contract and an expansive path. The contraction section is made of many contraction blocks. Each block takes an input applies two 3X3 convolution layers followed by a 2X2 max pooling. During the contraction, the spatial information is reduced while feature information is increased. The expansive path combines the feature and spatial information. Each block passes the input to two 3X3 CNN layers followed by a 2X2 upsampling layer. At the final layer a 1x1 convolution is used to map each 64-component feature vector to the desired number of classes. In total the network has 23 convolutional layers.

\section{Evaluation and metrics}

For the evaluation of the VGG-16 based scoring architecture, 5-fold cross-validation will be used to get a more realistic measure of the efficiency to correctly classify images with COVID-19 signs. And for the evaluation of the U-Net based segmentation architecture, the Intersection over Union (IoU) metric will be used, as it provides more accurate information on the degree of segmentation.

\section{Conclusions}

This paper presents the progress of the development of a portable system that makes use of lung ultrasound images in conjunction with two deep neural networks for the rapid detection of COVID-19. The final system will be implemented on a Raspberry Pi to perform inference. In this work we are also proposing a dataset which we are collecting and at its completion will be freely available. Although this project is interesting, we are still implementing the dataset.

An important aspect of this project is the use of ultrasound images. Thus, we use lung ultrasound because it does not use ionizing radiation, has low cost and has a high availability in health centers. Another advantage is that it can be used several times to evaluate the patient's condition, which is not possible with chest X-rays or CT scans. Another positive aspect of this project is its explainability, as it not only provides a prediction, but also shows the areas of why the prediction was made. 

At the end of this project, a positive contribution will be made and especially to people living in remote areas, since it will provide a tool for the diagnosis of COVID-19 and the follow-up of COVID-19 patients in their recovery.

\clearpage

{\small
\bibliographystyle{ieee_fullname}
\bibliography{references}
}

\end{document}